\documentclass[conference]{IEEEtran}
\IEEEoverridecommandlockouts

\usepackage{cite}
\usepackage{amsmath,amssymb,amsfonts}
\usepackage{algorithmic}
\usepackage{graphicx}
\usepackage{textcomp}
\usepackage{xcolor}
\def\BibTeX{{\rm B\kern-.05em{\sc i\kern-.025em b}\kern-.08em
    T\kern-.1667em\lower.7ex\hbox{E}\kern-.125emX}}

\usepackage{amsthm}

\newtheorem{proposition}{Proposition}[]
\newtheorem{theorem}{Theorem}[]

\theoremstyle{remark}

\begin{document}

\title{An Energy-efficient Ordered Transmission-based Sequential Estimation}

\author{\IEEEauthorblockN{Chen Quan, Geethu Joseph, and Nitin Jonathan Myers}
\IEEEauthorblockA{
\textit{Delft University of Technology}
\\
\textit{Emails: \{c.quan, g.joseph, n.j.myers\}@tudelft.nl}}
\vspace{-2mm}
}

\maketitle

\begin{abstract}
Estimation problems in wireless sensor networks typically involve gathering and processing data from distributed sensors to infer the state of an environment at the fusion center. However, not all measurements contribute significantly to improving estimation accuracy. The ordered transmission protocol, a promising approach for enhancing energy efficiency in wireless networks, allows for the selection of measurements from different sensors to ensure the desired estimation quality. In this work, we use the idea of ordered transmission to reduce the number of transmissions required for sequential estimation within a network, thereby achieving energy-efficient estimation. We derive a new stopping rule that minimizes the number of transmissions while maintaining estimation accuracy similar to general sequential estimation with unordered transmissions. Moreover, we derive the expected number of transmissions required for both general sequential estimation with unordered transmissions and proposed sequential estimation with ordered transmissions and make a comparison between the two systems. Simulation results indicate that our proposed scheme can efficiently reduce transmissions while still ensuring the quality of estimation.
\end{abstract}


\IEEEpeerreviewmaketitle

\vspace{-3mm}
\section{Introduction}

The ordered transmission protocol is a promising strategy for enhancing energy efficiency in wireless sensor networks, which is critical to extending the lifetime of sensor nodes \cite{feng2012survey}. The conventional ordered transmission protocol for distributed networks was first introduced in \cite{blum2008energy}. Here, only informative sensors transmitted their log-likelihood ratios (LLRs) to the fusion center (FC). This was achieved by a protocol that lets sensors transmit in decreasing order of the LLR magnitudes, with transmission start times proportional to $\frac{1}{|\text{LLR}|}$. Once the FC receives enough measurements to achieve the desired decision quality, it broadcasts a stop signal to halt further transmissions. This protocol has been applied to various problems, including noncoherent signal detection \cite{rawas2011energy}, sequential detection for cooperative spectrum sensing \cite{hesham2012distributed}, quickest change detection with independent \cite{chen2020optimal} and dependent measurements \cite{chen2021ordering}, as well as distributed training of machine learning models~\cite{chen2020ordered}. 
\par In quickest change detection, ordered transmission protocols can reduce the amount of information transferred without increasing detection delays, such as the worst-case average detection delay. In machine learning model training, these protocols can save bandwidth while achieving similar convergence rates as gradient descent for non-convex smooth loss functions. Additionally, in \cite{gupta2020ordered}, the ordered transmission scheme was integrated with energy harvesting to enhance the energy efficiency of wireless sensor networks. Spatial correlation among sensors was leveraged in \cite{gupta2020ordered2} to develop an ordered transmission scheme. In \cite{sriranga2018energy}, the authors devised a communication-efficient ordered transmission scheme in which informative sensors send binary decisions to the FC instead of LLRs, further enhancing the communication efficiency of the network. The aforementioned works demonstrate that ordered transmission-based schemes can effectively reduce the number of transmissions required for detection while maintaining the same inference performance. The authors in \cite{10251458,10043779} examined the security aspects of the ordered transmission protocol. They investigated the resilience of both conventional ordered transmission-based and communication-efficient ordered transmission-based systems under various Byzantine attacks, including mean-shift, mean-variance-shift \cite{10251458}, and data-flipping Byzantine attacks \cite{10043779}. 

Ordered transmission protocols are not limited to detection problems but have also been extended to certain estimation problems. The authors in \cite{5946994,5714758,5464939} designed ordered transmission-based estimation schemes under specific assumptions. In their work, it was assumed that the unknown parameters belong to a finite set of possible values. This allows each sensor to compute the LLRs for all possible parameter values and transmit them to the FC. By ordering these LLRs, the number of transmissions needed for global parameter estimation can be significantly reduced without any loss of performance. This approach works well when the unknown parameter lies within a discretized space, as calculating the LLRs for all possible values is usually straightforward. However, when dealing with a parameter in a continuous space, the methods in \cite{5946994,5714758,5464939} becomes impractical due to the infinite number of possible values. 

In this paper, we focus on applying the ordered transmission protocol to estimation problems where the unknown parameter lies within a continuous space. We assume that the continuous prior distribution for the unknown parameter is known, and that the Bayesian estimator is employed at the FC to estimate the parameter of interest. Since the overall uncertainty of the current estimate is influenced by the posterior variance, which, in turn, is influenced by the variance of the received measurements, we propose a new ordering strategy: ordering measurements based on their uncertainty. This approach allows us to focus on measurements that significantly contribute to estimation quality by prioritizing the transmission of those with lower uncertainty.

Our main contribution is the development of an ordered transmission-based sequential estimation approach for estimating unknown parameters in a continuous space. To assess the performance of our proposed scheme, we derive the expected number of transmissions in the ordered transmission-based framework for parameter estimation and compare it with the non-ordered case. Finally, through simulations, we demonstrate that our scheme significantly reduces the number of transmissions required in the network to solve a sequential estimation problem.

\section{System Model}
Consider a wireless sensor network consisting of $N$ sensor nodes distributed randomly over a region of interest. Each sensor node $i$ is equipped with a sensor to measure an unknown physical quantity $\theta$ (e.g., temperature, humidity) and a wireless module to communicate with a central FC. Each sensor node $i$ takes a noisy measurement of the true underlying parameter $\theta$, which follows a prior distribution $p(\theta)$. Here, we assume that the prior distribution for parameter $\theta$ follows Gaussian distribution with mean $\mu_0$ and variance $\tau_0^2$, i.e., $\theta \sim \mathcal{N}(\mu_0, \tau_0^2)$. The measurement $y_i$ at node $i$ is given by
\begin{equation}
y_i = \theta + w_i,
\end{equation}
where $w_i$ is the measurement noise, modeled as zero-mean Gaussian noise with variance $\sigma_i^2$, i.e., $w_i \sim \mathcal{N}(0, \sigma_i^2)$. Additionally, for the sake of analytical simplicity, we assume that the precision associated with its noise variance, i.e., $\frac{1}{\sigma_i^2}$, follows a uniform distribution over $[a,b]$.\footnote{Although we focus on a specific case where $w_i \sim \mathcal{N}(0, \sigma_i^2)$ and $\frac{1}{\sigma_i^2}\sim U(a,b)$, the proposed scheme is applicable to other distributions as well. This study, however, is beyond the scope of this paper.}

If the FC receives measurements from all the $N$ sensors, it could employ a Bayesian technique to estimate the unknown $\theta$. The optimal estimate of $\theta$ is its posterior mean, while its uncertainty is the posterior variance, both given by
\begin{align}
E[\theta \mid \mathbf{y}]\!=\! \frac{\frac{\mu_0}{\tau_0^2} \!+\! \sum_{i=1}^N \frac{y_i}{\sigma_i^2}}{\frac{1}{\tau_0^2} \!+\! \sum_{i=1}^N\frac{1}{\sigma_i^2}}, \quad
\text{Var}(\theta \mid \mathbf{y})\!=\!\frac{1}{\frac{1}{\tau_0^2} \!+\! \sum_{i=1}^N\frac{1}{\sigma_i^2}}.
\end{align}
Although this scheme provides optimal estimation, transmitting all measurements to the FC requires significant power and bandwidth resources. To address this limitation, we propose more efficient estimation schemes that reduce overhead.

We discuss two estimation schemes: the non-ordered sequential estimation scheme and the ordered sequential estimation scheme. In both schemes, not all sensors are required to send their measurements to the FC. The key difference is that the non-ordered sequential estimation scheme processes measurements in random order, whereas the ordered sequential estimation scheme prioritizes measurements with lower uncertainty. We next examine the average number of transmissions required for both these schemes to achieve the same desired accuracy.

\subsection{Non-ordered Sequential Estimation Scheme} \label{sec:unordered}
In this scheme, the sensor measurements are collected sequentially and in a random order, which allows a sequential update of the unknown parameter $\theta$. When enough data has been collected to make a reliable estimate, the transmission stops. A common stopping rule in sequential estimation involves setting a threshold on the width of the confidence interval. The confidence intervals can be updated after each measurement, and stopping rules can be established based on the width of the desired confidence interval. The estimate of $\theta$ and its variance after observing $k \leq N$ measurements are given by
\begin{equation}
    \hat{\theta}_{\mathrm{FC}}=E[\theta \mid \{y_i\}_{i=1}^k]=\frac{\frac{\mu_0}{\tau_0^2} + \sum_{i=1}^k \frac{y_i}{\sigma_i^2}}{\frac{1}{\tau_0^2} + \sum_{i=1}^k\frac{1}{\sigma_i^2} },
\end{equation}
and its variance is $\text{Var}(\theta \mid \{y_i\}_{i=1}^k)={\left(\frac{1}{\tau_0^2} + \sum_{i=1}^k\frac{1}{\sigma_i^2}\right)}^{-1}$. The confidence interval for $\theta\mid \{y_i\}_{i=1}^k$ is defined as
\begin{align}\label{eq:confi_interv}
    \text{CI} \!=\! &\left( \!\hat{\theta}_{\mathrm{FC}} \!-\! \alpha \sqrt{\text{Var}(\theta \!\mid\!\! \{y_i\}_{i=1}^k)}, \hat{\theta}_{\mathrm{FC}} \!+\! \alpha  \sqrt{\text{Var}(\theta \!\mid\!\! \{y_i\}_{i=1}^k}) \!\right),
\end{align}
which is $\alpha$ standard deviations on either side of the mean. Here, $\alpha$ is a constant that determines the width of the confidence interval. Due to the Gaussian posterior in our problem, it can be shown that $\alpha$ is approximately 1.96 to achieve a $95\%$ confidence interval. The confidence interval updates after each measurement, and the updating stops when the width of the confidence interval is below a specified threshold $\epsilon$, i.e.,
\begin{align}\label{eq:stoprule_se}
    2\alpha \sqrt{\text{Var}(\theta \mid \{y_i\}_{i=1}^k)}\leq 2 \epsilon 
\end{align}
This ensures that the estimate achieves a certain level of accuracy. By monitoring the width of the confidence interval, the FC can determine when to stop collecting new measurements. Thus, according to \eqref{eq:stoprule_se}, we can derive the stopping rule for sequential estimation as follows:
\begin{equation}\label{eq:unoreder_stop}
    \sum_{i=1}^kz_i \geq \left(\frac{\alpha}{\epsilon}\right)^2-\frac{1}{\tau_0^2}=\gamma,
\end{equation}
where $z_i=\frac{1}{\sigma_i^2}\geq 0$. The remaining $N-k$ sensors in the network stop transmissions when the FC receives $k$ measurements that meet the stopping criterion in \eqref{eq:unoreder_stop}. To obtain the average number of transmissions needed to achieve a desired confidence interval, we need to compute the expected number of transmissions $E[k^*]$, where  $k^*$ represents the minimum number of measurements needed to satisfy \eqref{eq:unoreder_stop}. 

\begin{proposition}\label{pro:1}
For non-ordered sequential estimation scheme with stopping rule~\eqref{eq:unoreder_stop}, the expected number of transmissions $E[k^*]$ is given as 
\begin{align}
    E[k^*]=\sum_{k=0}^{N-1}\mathrm{Pr}(\sum_{i=1}^{k}z_i\leq\gamma)\label{eq:expected_N_t3}.
\end{align}
\end{proposition}
\begin{IEEEproof}
    See Appendix \ref{proof_pro:1}.
\end{IEEEproof}
According to Proposition~\ref{pro:1}, to compute $E[k^*]$, we need to first compute $\mathrm{Pr}(\sum_{i=1}^{k}z_i\leq\gamma)$. Let $S_k=\sum_{i=1}^{k}z_i$ and $S_k^{'}=\sum_{i=1}^{k}z_i^{'}$, where $z_i^{'}=\frac{z_i-a}{b-a}$. Clearly, $z_i^{'} \sim U(0,1)$ for $i\in\{1,\dots,N\}$, meaning that the normalized variables $z_i^{'}$ are uniformly distributed on $[0,1]$. By normalizing $z_i$, we can leverage properties of sums of uniformly distributed random variables to compute $\mathrm{Pr}(\sum_{i=1}^{k}z_i\leq\gamma)$. We have
\begin{align}\label{eq:transformation_unorder}
    \mathrm{Pr}(\sum_{i=1}^{k}z_i\leq\gamma)=&\mathrm{Pr}(S_k^{'}\leq\gamma^{'}),
\end{align}
where $\gamma^{'}=\frac{\gamma-ka}{b-a}$. Substituting \eqref{eq:transformation_unorder} back into \eqref{eq:expected_N_t3}, the expected number of transmissions
required $E[k^*]$ is given by
\begin{equation}\label{eq:unordered_en}
    E[k^*]=\sum_{k=0}^{N-1} \frac{1}{k!}\sum_{j=0}^{\gamma^{*}} (-1)^j {k\choose j}(\min(\gamma^{'},k)\!-\!j)^k,
\end{equation}
where $\gamma^{*}=\min(\max(\lfloor\gamma^{'}\rfloor,0),k)$ and $\lfloor\gamma^{'}\rfloor$ denotes the floor of $\gamma^{'}$ to the next lowest integer. The expression in \eqref{eq:unordered_en} is derived using the fact that the sum of independent uniformly distributed random variables on $[0,1]$ follows the Irwin-Hall distribution.

\subsection{Ordered Transmission Scheme Model}
In the non-ordered sequential estimation scheme discussed in Sec. \ref{sec:unordered}, the FC receives one measurement at a time in a random order. In contrast, the ordered sequential estimation scheme prioritizes measurements with lower statistical uncertainty.
Specifically, the uncertainty in the current estimate of $\theta$ at the FC is characterized by the posterior variance $\text{Var}(\theta \mid \{y_i\}_{i=1}^k)$, which is a function of $\sum_{i=1}^k (1/{\sigma^2_i})$. Thus, the uncertainty in the current estimate of $\theta$ is affected by the uncertainty in the individual sensor measurements. To optimize the estimation process, it is advantageous to let the sensors send their measurements to the FC in the ascending order of the individual sensor uncertainties, i.e., $\{\sigma_i^2\}_{i=1}^N$. To achieve this goal, sensor $i$ can be equipped with a synchronized timer, with the start time set proportionally to the variance $\sigma_i^2$. As the timer counts down to zero, the sensor transmits its measurement to the FC. This approach prioritizes sensors with lower uncertainty, facilitating faster convergence to an estimate with the desired accuracy.

To represent the ordered transmissions, we arrange $\{1/ \sigma^2_i\}_{i=1}^N$ in the descending order. This sequence is denoted by $z_{[1]}>z_{[2]}>\ldots>z_{[N]}$, where $z_{[i]}$ indicates the $i^{\text{th}}$ largest $z$, i.e., the $i^{\text{th}}$ largest value in the descending sequence of inverted variances. With this sequence, the sensor with the precision $z_{[1]}$ transmits its corresponding measurement $y_{[1]}$ first, followed by the sensor with $z_{[2]}$, which transmits $y_{[2]}$, and so on. Let $k$ denote the number of measurements received by the FC. Again, the FC can estimate $\theta$ using a Bayesian approach. The estimate of the unknown parameter is given by 
\begin{align}
\hat{\theta}^{OT}_{FC}&=E[\theta\mid\{y_{[i]}\}_{i=1}^k]=\frac{\frac{\mu_0}{\tau_0^2} + \sum_{i=1}^k \frac{y_{[i]}}{\sigma_{[i]}^2}}{\frac{1}{\tau_0^2} + \sum_{i=1}^k\frac{1}{\sigma_{[i]}^2}}
\end{align} 
Following a similar procedure from \eqref{eq:confi_interv} to \eqref{eq:unoreder_stop}, we can derive the stopping rule for ordered sequential estimation given a desired confidence interval $\epsilon$. The stopping rule is expressed as
\begin{equation}\label{eq:order_stop}
    \sum_{i=1}^kz_{[i]} \geq \left(\frac{\alpha}{\epsilon}\right)^2-\frac{1}{\tau_0^2}=\gamma,
\end{equation}
where $z_{[i]}=\frac{1}{\sigma^2_{[i]}}$. The expected number of measurements required under the proposed stopping rule is discussed in Theorem \ref{thm:ub_lb2}.

\begin{theorem}
\label{thm:ub_lb2}
For ordered sequential estimation, the expected number of measurements required $E[k^*]$ can be bounded as $\bar{N}_{t}^L\leq E[k^*]\leq \bar{N}_{t}^U$. Here, the upper bound $ \bar{N}_{t}^U$ and the lower bound $\bar{N}_{t}^L$ are given by 
\begin{align}\label{eq:ub}
    \bar{N}_{t}^U=\sum_{k=0}^{N-1}\frac{N!\left(\frac{1}{b-a}\right)^N}{(k!)^2(N-k-1)!}\sum_{j=0}^k(-1)^j{k\choose j}S_1,
\end{align}
\begin{align}\label{eq:lb}
    \bar{N}_{t}^L=\sum_{k=0}^{N-1}\left(\frac{\gamma/k-a}{b-a}\right)^N,
\end{align}
where $S_1$ is given in \eqref{eq:s1}.
\begin{figure*}[ht]
\begin{equation}\label{eq:s1}
    S_1=\left\{
    \begin{array}{lll}
    \sum_{i=0}^k{N-k-1\choose i}(\gamma-bj)^i(j-k)^{k-i}\!\left[\sum_{m=0}^{N-k-1}\!\!\!{N\!-\!k\!-\!1\choose m}\!(\!-\!a)^{N\!-\!k\!-\!1\!-\!m}\frac{c^{k\!-\!i\!+\!m\!+\!1}\!-\!a^{k\!-\!i\!+\!m\!+\!1}}{k\!-\!i\!+\!m\!+\!1}\right],&\text{if $\gamma\!<\! kb$},\\
    \sum_{i=0}^{N\!-\!k\!-\!1}\!\!{N-k-1\choose i}(k-j)^k(-a)^{N-k-1-i}\!\left[\sum_{m\!=\!0}^{k}{k\choose m}b^m(-1)^{k\!-\!m}\frac{b^{k\!-\!m\!+\!i\!+\!1}\!-\!a^{k\!-\!m\!+\!i\!+\!1}}{k-m\!+\!i\!+\!1}\right],&\text{if $\gamma\geq kb$.}\\
    \end{array}\right. 
\end{equation}
\vspace{-0.8cm}
\end{figure*}
\end{theorem}
\begin{IEEEproof}
See Appendix~\ref{thm:1}.
\end{IEEEproof}
The reason we seek upper and lower bounds instead of the exact expression for the expected number of measurements is that computing the exact expression is extremely computationally expensive (see ~\cite[Theorem 3.2]{10251458}).
\vspace{-1mm}
\section{Numerical and simulation results}

\begin{figure}[htbp]
\centerline{\includegraphics[height=6.5cm,width=9.5cm]{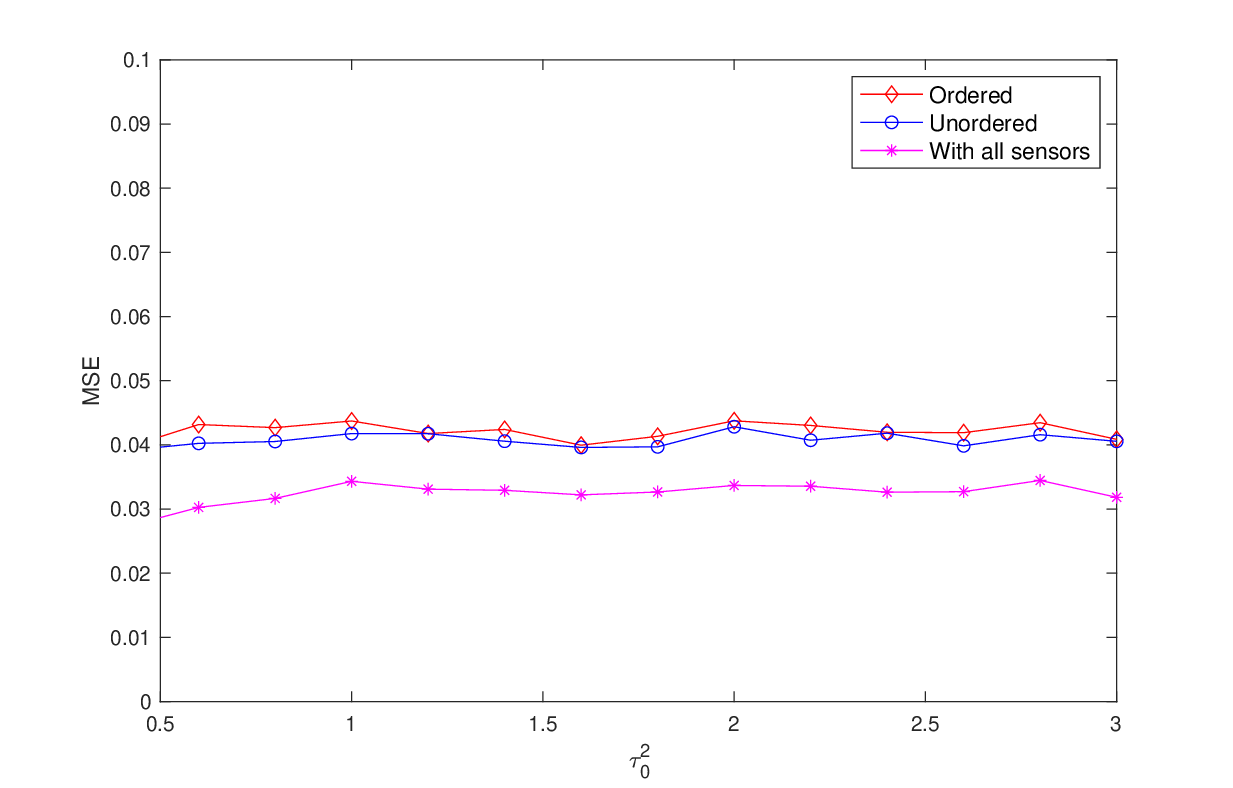}}
\caption{Comparison of MSE as a function of $\tau_0^2$ for ordered and unordered sequential estimation systems, as well as general estimation systems. For the ordered and unordered sequential estimation systems, a specified desired confidence interval is set with $\epsilon=0.4$.}
\label{fig:same_conf_interv}
\end{figure}

In this section, we present some numerical and simulation results to support our theoretical analysis. We assume $N=50$ sensors and a prior distribution with $\mu_0=2$. The noise variance distribution parameters are set to $a=1/5$ and $b=1$, while the confidence threshold is $\epsilon=0.4$. Fig. \ref{fig:same_conf_interv} shows the mean square error (MSE) in parameter estimation as a function of $\tau_0^2$ for systems that adopt the unordered sequential estimation scheme, the ordered sequential estimation scheme, and the system that utilizes all the sensor measurements. We can observe that the systems adopting the unordered and ordered sequential estimation schemes have nearly the same MSE values for the same confidence interval. Additionally, the system using measurements from all the sensors has the lowest MSE, as generally, the more measurements we take, the less uncertainty remains. 

\begin{figure}[htbp]
\centerline{\includegraphics[height=6.5cm,width=9.5cm]{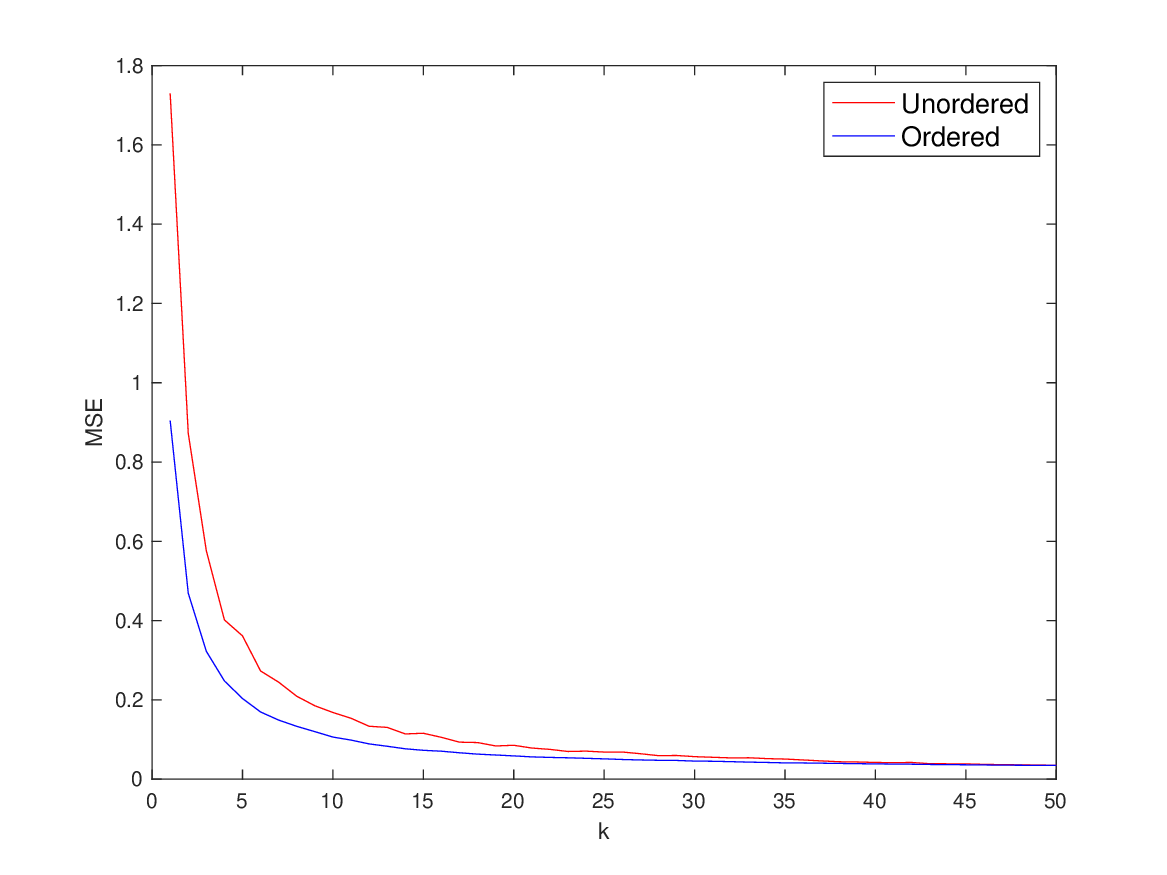}}
\caption{Comparison of MSE as a function of $k$ for ordered and unordered sequential estimation systems.}
\label{fig:fixed_n}
\end{figure}

Fig. \ref{fig:fixed_n} shows the MSE as a function of the number of measurements, $k$, for systems adopting the unordered sequential estimation scheme and the ordered sequential estimation scheme. We observe that the system adopting the ordered sequential estimation scheme has a smaller MSE compared to the unordered case. This is due to the fact that estimations based on measurements with statistically lower uncertainty generally result in more accurate estimates. In the system adopting the ordered sequential estimation scheme, the measurements with statistically lower uncertainty are prioritized and processed first. For the same number of measurements, the system that randomly selects measurements, as in the unordered sequential estimation scheme, does not outperform the ordered scheme.

\begin{figure}[htbp]
\centerline{\includegraphics[height=6.5cm,width=9.5cm]{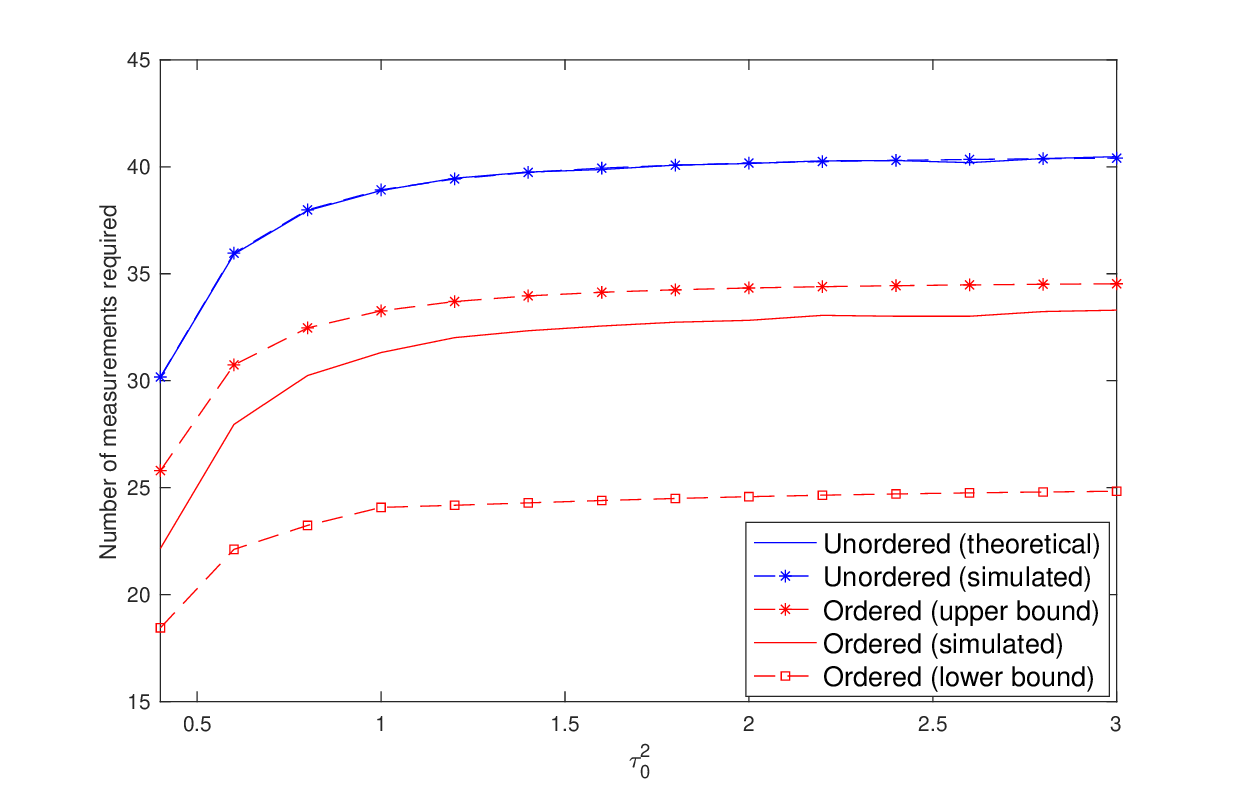}}
\caption{Comparison of the average number of measurements required as a function of $\tau_0^2$ for ordered and unordered sequential estimation systems for a specified desired confidence interval with $\epsilon=0.4$.}
\label{fig:En}
\end{figure}

Fig. \ref{fig:En} shows the MSE as a function of $\tau_0^2$ for systems that adopt the unordered sequential estimation scheme and the ordered sequential estimation scheme. We compare the results obtained via Monte Carlo simulations with our analytical results for both the unordered and ordered system. We can observe a good match between the analytical results from \eqref{eq:unordered_en} and the simulated results from the Monte Carlo method. The upper bound obtained from \eqref{eq:ub} and the lower bound from \eqref{eq:lb} can accurately track the variation in the underlying average number of transmissions required. The upper bound we obtained is relatively tight, effectively indicating the maximum number of measurements required. It can also be observed that the expected number of measurements required increases as $\tau_0^2$ increases. Since the thresholds in both \eqref{eq:unoreder_stop} and \eqref{eq:order_stop} increase as $\tau_0^2$ increases, the number of required measurements also increases. However, due to the characteristics of the $\frac{1}{\tau_0^2}$ term in the expressions for the thresholds, as $\tau_0^2$ increases, the rate of increase of $-1/{\tau_0^2}$ decreases. Consequently, the rate of increase in the number of required measurements decreases. Fig. \ref{fig:En} also demonstrates that the system adopting the ordered sequential estimation scheme can efficiently reduce the number of transmissions without any loss in estimation accuracy.

\vspace{-3mm}
\section{Conclusion}
\label{sec:conclusion}
In this paper, we proposed an ordered sequential estimation scheme that employs an ordered transmission protocol to minimize the number of transmissions required while maintaining a desired level of estimation accuracy. We derived the stopping rule for this scheme and calculated the expected number of measurements needed for both ordered and unordered sequential estimation methods. Our numerical and simulation results demonstrated that the ordered sequential estimation scheme can significantly reduce the number of transmissions without sacrificing estimation accuracy. This approach highlights how to design systems with lower communication overhead while preserving nearly the same estimation quality. In the future, we aim to investigate several extensions of our problem, including scenarios in which the variances of the measurements are unknown, cases where the FC only receives quantized versions of the measurements, and situations in which some falsified measurements are reported to the FC.

\appendices

\section{Proof of Proposition \ref{pro:1}}\label{proof_pro:1}
The expected number of transmissions required in the network is given as
\begin{subequations}\label{eq:expected_N_t}
\begin{align}
    E[k^*]&=\sum^N_{k=1}k\mathrm{Pr}(k^*=k)=\sum_{k=1}^{N}\mathrm{Pr}(k^*\geq k)\label{eq:expected_N_t2}\\
    &=N-\sum_{k=0}^{N-1}\mathrm{Pr}(k^*\leq k)\\
    &=N-\sum_{k=0}^{N-1}\mathrm{Pr}(\sum_{i=1}^{k}z_i\geq\gamma)\label{eq:expected_N_t1}\\
    &=\sum_{k=0}^{N-1}\mathrm{Pr}(\sum_{i=1}^{k}z_i\leq\gamma),
\end{align}
\end{subequations}
where $k^*$ is the minimum number of measurements required to satisfy the equality in \eqref{eq:unoreder_stop}. Here, \eqref{eq:expected_N_t2} can be derived by induction using the  observation that $\sum^N_{k=1}k\mathrm{Pr}(k^*=k)= \mathrm{Pr}(k^*=1)+2\mathrm{Pr}(k^*=2)+3\mathrm{Pr}(k^*=3)+\cdots +N \mathrm{Pr}(k^*=N) = \mathrm{Pr}(k^*\geq 1)+\mathrm{Pr}(k^*=2)+2\mathrm{Pr}(k^*=3)+\cdots +(N-1)\mathrm{Pr}(k^*=N)$. Since $\{z_i\}_{i=1}^N$ are non-negative, \eqref{eq:expected_N_t1} follows from the fact that if $\sum_{i=1}^{k^*}z_i\geq\gamma$, then $\sum_{i=1}^{k}z_i\geq\gamma$ for any $k\geq k^*$.

\section{Proof of Theorem \ref{thm:ub_lb2}}
\label{thm:1}
The expected number of measurements required for ordered sequential estimation is given by $E[k^*]=\sum_{k=0}^{N-1}\mathrm{Pr}(M_k\leq \gamma)$, following a similar procedure as in \eqref{eq:expected_N_t}, where $M_k=\sum_{i=1}^{k}z_{[i]}$. The difference here is that $k^*$ now is the minimum number of measurements required to satisfy the equality in \eqref{eq:order_stop}. To compute $E[k^*]$, we need to first compute $\mathrm{Pr}(M_k\leq\gamma)$ and it can be expressed as
\begin{align}\label{eq:int}
    \!\!\mathrm{Pr}(M_k\!\leq\!\gamma)\!=\!\!\int_{a}^{b}\!\!\!\mathrm{Pr}(M_k\!\leq\!\gamma|z_{[k+1]}\!=\!x)f(z_{[k+1]}\!=\!x)dx,
\end{align}
where the probability density 
\begin{align}\label{eq:ordered_statistic}
    f(z_{[k+1]}\!=\!x)=\frac{N!\left(\frac{1}{b\!-\!a}\right)^N}{(N\!-\!k\!-\!1)!k!}\left(x\!-\!a\right)^{N\!-\!k\!-\!1}\left(b\!-\!x\right)^{k}.
\end{align}
Next, we need to compute $\mathrm{Pr}(M_k\!\leq\!\gamma|z_{[k+1]}\!=\!x)$. However, obtaining an exact expression for this probability is non-trivial and computationally expensive (see ~\cite[Theorem 3.2]{10251458}). Nevertheless, we can obtain an upper bound on the expected number of measurements by performing a similar transformation as in \eqref{eq:transformation_unorder}.
Let $M_k^{'}=\sum_{i=1}^{k}z_{[i]}^{'}$, where $z_{[i]}^{'}=\frac{z_{[i]}-x}{b-x}$. Due to the fact that $z_{[i]}^{'} \in [0,1]$ for $i\in\{1,\dots,k\}$ and $z_{[1]}\geq\dots\geq z_{[k+1]}$, $\mathrm{Pr}(M_k\leq\gamma|z_{[k+1]}=x)$ is given by
\begin{subequations}
    \begin{align}
    &\mathrm{Pr}(M_k\leq\gamma|z_{[k+1]}=x)\notag\\
    &\;\;=\mathrm{Pr}(M_k\leq\gamma,z_{[1]}\geq x,\dots,z_{[k]}\geq x|z_{[k+1]}=x)\\
    &\;\;= \mathrm{Pr}\left(M_k^{'}\!\leq\!\frac{\gamma\!-\!kx}{b\!-\!x},z_{[1]}\!\geq\! x,\dots,z_{[k]}\!\geq\! x|z_{[k+1]}\!=\!x\right)\\
    &\leq \mathrm{Pr}\left(M_k^{'}\leq\frac{\gamma-kx}{b-x}|z_{[k+1]}=x\right)\label{eq:ub_derive}\\
    &=\frac{1}{k!} \sum_{j=0}^{k} (-1)^j {k\choose j}(f(x)-j)_{+}^k,\label{eq:condi_dis}
    \end{align}
\end{subequations}
where we define
\begin{equation}
    (f(x) - j)_{+} = 
\begin{cases}
f(x) - j & \text{if } f(x) - j \geq 0 \\
0 & \text{if } f(x) - j < 0,
\end{cases}
\end{equation}
and
$f(x)=\max(\min(\frac{\gamma-kx}{b-x},k),0)$. Here, \eqref{eq:ub_derive} follows from the fact that the probability of the intersection of events can never exceed the probability of any one of the individual events. 
Also, \eqref{eq:condi_dis} follows from the fact that if $z_{[1]}^{'},\dots,z_{[k]}^{'} \in [0, 1]$, the sum of $z_{[1]}^{'},\dots,z_{[k]}^{'}$ is independent of their order. Thus, $M_k^{'}$ can be treated as the sum of $k$ independent uniformly distributed random variables on $[0,1]$, which follows the Irwin-Hall distribution and is easy to compute. Now, by substituting the expressions from \eqref{eq:condi_dis} and \eqref{eq:ordered_statistic} into \eqref{eq:int}, we can derive the upper bound of $P(M_k\leq\gamma)$ as
    \begin{multline}
    P(M_k\leq\gamma)\leq\frac{N!}{(N\!-\!k\!-\!1)!(k!)^2}\left(\frac{1}{b-a}\!\right)^N\sum_{j=0}^{k} (-1)^j {k\choose j}\\
    \times\int_{a}^{b}g(x)^k(x-a)^{N-k-1}(b-x)^kdx,\label{eq:g(x)}
\end{multline}
where $g(x)=(f(x)-j)_{+}=\mathrm{max}((f(x)-j),0)$, and it can be further deduced that
\begin{align}\label{eq:g(x)1}
g(x) = \begin{cases}
k - j & \text{if } \frac{\gamma - kx}{b - x} \geq k \text{ and } k > j \\
\frac{\gamma - kx}{b - x} - j & \text{if } 0\leq\frac{\gamma - kx}{b - x} < k \text{ and } \frac{\gamma - kx}{b - x} > j \\
0 & \text{otherwise}.
\end{cases}
\end{align}
Since $x\in[a,b]$ and $j\leq k$, we can write \eqref{eq:g(x)1} as 
\begin{align}\label{eq:g(x)2}
g(x) = \begin{cases}
k - j & \text{if } \gamma\geq kb\\
\frac{\gamma - kx}{b - x} - j & \text{if } \gamma< kb \text{ and } a\leq x\leq c \\
0 & \text{otherwise},
\end{cases}
\end{align}
where $c=\min\left(\frac{\gamma - jb}{k-j},\frac{\gamma}{k}\right)$.
Thus, we need to consider two cases when we compute the term $\int_{a}^{b}g(x)^k(x-a)^{N-k-1}(b-x)^kdx$.

When $\gamma< kb$, the term $\int_{a}^{b}g(x)^k(x-a)^{N-k-1}(b-x)^kdx$ in \eqref{eq:g(x)} is given by
\begin{subequations}
    \begin{align}
    &\int_{a}^{b}g(x)^k(x-a)^{N-k-1}(b-x)^kdx\notag\\
    &\;\;=\int_{a}^{c}[(\gamma-bj)+(j-k)x]^k(x-a)^{N-k-1}dx\\
    &\;\;=\int_{a}^{c}\!\!\left[\sum_{i=0}^k\!\!{k\choose i}(\gamma\!-\!bj)^i(j\!-\!k)^{k\!-\!i}x^{k\!-\!i}\right]\!\!(x\!-\!a)^{N\!-\!k\!-\!1}dx\\
    &\;\;=\sum_{i=0}^k\!{k\choose i}(\gamma\!-\!bj)^i(j\!-\!k)^{k\!-\!i}\int_{a}^{c}x^{k\!-\!i}(x\!-\!a)^{N\!-\!k\!-\!1}dx\\
    &\;\;=\sum_{i=0}^k{k\choose i}(\gamma-bj)^i(j-k)^{k-i}\notag\\
    &\quad\times\left[\!\sum_{m=0}^{N\!-\!k\!-\!1}\!\!\!{N\!-\!k\!-\!1\choose m}(-a)^{N\!-\!k\!-\!1\!-\!m}\int_{a}^{c}x^{k\!-\!i\!+\!m}dx\right]\\
    &\;\;=\sum_{i=0}^k{N-k-1\choose i}(\gamma-bj)^i(j-k)^{k-i}\!
\notag\\
    &\quad\left[\!\sum_{m=0}^{N-k-1}\!\!{N\!-\!k\!-\!1\choose m}\!(\!-\!a)^{N\!-\!k\!-\!1\!-\!m}\frac{c^{k-i+m+1}\!-\!a^{k-i+m+1}}{k\!-\!i\!+\!m\!+\!1}\right],\label{ub:1}
\end{align}
\end{subequations}
according to Binomial Theorem.

When $\gamma\geq kb$, $\int_{a}^{b}g(x)^k(x-a)^{N-k-1}(b-x)^kdx$ in \eqref{eq:g(x)} is given by
\begin{subequations}
    \begin{align}
        &\int_{a}^{b}g(x)^k(x-a)^{N-k-1}(b-x)^kdx\\
    &\;\;=(k-j)^k\int_{a}^{b}(x-a)^{N-k-1}(b-x)^kdx\\
    &\;\;=(k\!-\!j)^k\!\!\!\int_{a}^{b}\!\!\left[\sum_{i=0}^{N\!-\!k\!-\!1}\!\!\!{N\!-\!k\!-\!1\choose i}x^i(\!-\!a)^{N\!-\!k\!-\!1\!-\!i}\right]\!\!(b\!-\!x)^{k}dx\\
    &\;\;=\sum_{i=0}^{N\!-\!k\!-\!1}\!\!\!{N\!-\!k\!-\!1\choose i}(k\!-\!j)^k(\!-\!a)^{N\!-k\!-\!1-i}\int_{a}^{b}\!\!x^i(b\!-\!x)^{k}dx\\
    &\;\;=\sum_{i=0}^{N-k-1}{N-k-1\choose i}(k-j)^k(-a)^{N-k-1-i}\notag\\
    &\quad\times\int_{a}^{b}\left[\sum_{m=0}^{k}{k\choose m}b^m(-1)^{k-m}x^{k-m+i}\right]dx\\
    &\;\;=\sum_{i=0}^{N-k-1}{N-k-1\choose i}(k-j)^k(-a)^{N-k-1-i}\notag\\
    &\quad\times\left[\!\sum_{m=0}^{k}\!\!{k\choose m}b^m(\!-\!1)^{k-m}\frac{b^{k-m+i+1}\!-\!a^{k-m+i+1}}{k-m+i+1}\!\right].\label{ub:2}
    \end{align}
\end{subequations}
Substituting the expressions from \eqref{ub:1} and \eqref{ub:2} into \eqref{eq:g(x)}, we can obtain the final expression of upper bound as shown in Theorem~\ref{thm:ub_lb2}. Then, we compute one possible lower bound for $\mathrm{Pr}(M_k\leq\gamma)$ and it is given by
\begin{subequations}
    \begin{align}
    \label{eq:lb_the_st1}
    \mathrm{Pr}(M_k\leq\gamma)
    &\geq \mathrm{Pr}(z_{[1]}\leq \frac{\gamma}{k},z_{[2]}\leq \frac{\gamma}{k},\dots,z_{[k]}\leq \frac{\gamma}{k})\\
    \label{eq:lb_the_st2}
    &\;\;=\mathrm{Pr}(z_{1}\leq \frac{\gamma}{k},z_{2}\leq \frac{\gamma}{k},\dots,z_{N}\leq \frac{\gamma}{k})\\
    &\;\;=\left(\frac{\gamma/k-a}{b-a}\right)^N\label{eq:lb_th}.
\end{align}
\end{subequations}
Here, \eqref{eq:lb_the_st1} is due to the fact that  $z_{[1]}\leq{\gamma}/{k},z_{[2]}\leq {\gamma}/{k},\dots,z_{[k]}\leq {\gamma}/{k}$ is only a subset of the feasible combinations of $z_{[1]}, z_{[2]}, \cdots z_{[k]}$ that sum to a value no larger than $\gamma$. Further, \eqref{eq:lb_the_st1} follows from the observation that all the $z_i$s need to be upper bounded by $\gamma/k$ when it is known that $z_{[1]} \leq \gamma/k$. Now, we obtain the final expression of the lower bound shown in Theorem~\ref{thm:ub_lb2}.

\bibliography{refer.bib}

\begin{thebibliography}{10}
\providecommand{\url}[1]{#1}
\csname url@samestyle\endcsname
\providecommand{\newblock}{\relax}
\providecommand{\bibinfo}[2]{#2}
\providecommand{\BIBentrySTDinterwordspacing}{\spaceskip=0pt\relax}
\providecommand{\BIBentryALTinterwordstretchfactor}{4}
\providecommand{\BIBentryALTinterwordspacing}{\spaceskip=\fontdimen2\font plus
\BIBentryALTinterwordstretchfactor\fontdimen3\font minus \fontdimen4\font\relax}
\providecommand{\BIBforeignlanguage}[2]{{%
\expandafter\ifx\csname l@#1\endcsname\relax
\typeout{** WARNING: IEEEtran.bst: No hyphenation pattern has been}%
\typeout{** loaded for the language `#1'. Using the pattern for}%
\typeout{** the default language instead.}%
\else
\language=\csname l@#1\endcsname
\fi
#2}}
\providecommand{\BIBdecl}{\relax}
\BIBdecl

\bibitem{feng2012survey}
D.~Feng, C.~Jiang, G.~Lim, L.~J. Cimini, G.~Feng, and G.~Y. Li, ``A survey of energy-efficient wireless communications,'' \emph{IEEE Commun. Surv. Tutorials}, vol.~15, no.~1, pp. 167--178, 2012.

\bibitem{blum2008energy}
R.~S. Blum and B.~M. Sadler, ``Energy efficient signal detection in sensor networks using ordered transmissions,'' \emph{IEEE Trans. Signal Process.}, vol.~56, no.~7, pp. 3229--3235, 2008.

\bibitem{rawas2011energy}
Z.~N. Rawas, Q.~He, and R.~S. Blum, ``Energy-efficient noncoherent signal detection for networked sensors using ordered transmissions,'' in \emph{Proc. Annu. Conf. Inf. Sci. Syst.}, 2011, pp. 1--5.

\bibitem{hesham2012distributed}
L.~Hesham, A.~Sultan, M.~Nafie, and F.~Digham, ``Distributed spectrum sensing with sequential ordered transmissions to a cognitive fusion center,'' \emph{IEEE Trans. Signal Process.}, vol.~60, no.~5, pp. 2524--2538, 2012.

\bibitem{chen2020optimal}
Y.~Chen, R.~S. Blum, and B.~M. Sadler, ``Optimal quickest change detection in sensor networks using ordered transmissions,'' in \emph{Proc. IEEE Int. Work. Signal Process. Adv. Wireless Commun.}, 2020, pp. 1--5.

\bibitem{chen2021ordering}
------, ``Ordering for communication-efficient quickest change detection in a decomposable graphical model,'' \emph{IEEE Trans. Signal Process.}, vol.~69, pp. 4710--4723, 2021.

\bibitem{chen2020ordered}
Y.~Chen, B.~M. Sadler, and R.~S. Blum, ``Ordered gradient approach for communication-efficient distributed learning,'' in \emph{Proc. IEEE Int. Work. Signal Process. Adv. Wireless Commun.}, 2020, pp. 1--5.

\bibitem{gupta2020ordered}
S.~S. Gupta, S.~K. Pallapothu, and N.~B. Mehta, ``Ordered transmissions for energy-efficient detection in energy harvesting wireless sensor networks,'' \emph{IEEE Trans. Commun.}, vol.~68, no.~4, pp. 2525--2537, 2020.

\bibitem{gupta2020ordered2}
S.~S. Gupta and N.~B. Mehta, ``Ordered transmissions schemes for detection in spatially correlated wireless sensor networks,'' \emph{IEEE Trans. Commun.}, vol.~69, no.~3, pp. 1565--1577, 2020.

\bibitem{sriranga2018energy}
N.~Sriranga, K.~G. Nagananda, R.~S. Blum, A.~Saucan, and P.~K. Varshney, ``Energy-efficient decision fusion for distributed detection in wireless sensor networks,'' in \emph{Proc. Int. Conf. Inf. Fusion}, 2018, pp. 1541--1547.

\bibitem{10251458}
C.~Quan, S.~Bulusu, B.~Geng, Y.~S. Han, N.~Sriranga, and P.~K. Varshney, ``On ordered transmission based distributed gaussian shift-in-mean detection under byzantine attacks,'' \emph{IEEE Trans. Signal Process.}, vol.~71, pp. 3343--3356, 2023.

\bibitem{10043779}
C.~Quan, N.~Sriranga, H.~Yang, Y.~S. Han, B.~Geng, and P.~K. Varshney, ``Efficient ordered-transmission based distributed detection under data falsification attacks,'' \emph{IEEE Signal Process. Lett.}, vol.~30, pp. 145--149, 2023.

\bibitem{5946994}
R.~S. Blum, ``Ordering for energy efficient estimation and optimization in sensor networks,'' in \emph{Proc. IEEE Int. Conf. Acoust. Speech Signal Process.}, 2011, pp. 2508--2511.

\bibitem{5714758}
------, ``Ordering for estimation and optimization in energy efficient sensor networks,'' \emph{IEEE Trans. Signal Process.}, vol.~59, no.~6, pp. 2847--2856, 2011.

\bibitem{5464939}
------, ``Ordering for estimation,'' in \emph{Proc. Annu. Conf. Inf. Sci. Syst.}, 2010, pp. 1--6.

\end{thebibliography}
\bibliographystyle{IEEEtran}

\end{document}